\documentclass[graphicx,epsfig,floats]{revtex4}
\usepackage{longtable}
\usepackage{epsfig}
\usepackage{amsmath}
\usepackage{amsfonts}
\usepackage{graphicx}
\usepackage{amssymb}
\usepackage{amsbsy}
\usepackage{tabularx}
\usepackage[autostyle]{csquotes}

\begin{document}

\title {High resolution spectroscopy on Te$_2$: new lines for reference}

\author{T.~Dutta$^1$ and M.~Mukherjee$^{1,2}$}
\affiliation{$^1$Centre for Quantum Technologies, National
University Singapore, Singapore 117543}
\affiliation{$^2$Department of Physics, National University Singapore, Singapore 117551. }

\date{\today}

\begin{abstract} 

 Ro-vibrational spectra of different electronic states of molecules are often used as absolute wavelength or frequency standards. These standards are also used to mitigate any slow drift of laser frequency during an experiment. In precision experiment, the two most commonly used molecular standards are iodine and tellurium, both are homo-nuclear diatomic molecules. The former is mostly used as standard for the long wavelength ($600-900$~nm) region, while the tellurium spectrum is widely used in short wavelength ($400-550$~nm) including near ultra violet. A comprehensive data on tellurium spectra can be obtained from the tellurium atlas~\cite{Te2atlas:80}. However near the $455~$nm range where a number of important atomic resonance line, the atlas provides no significant data. We have performed high resolution modulation transfer spectroscopy~(MTS) on tellurium molecule in a hot cell in the region close to $455~$nm wavelength thereby obtained more than $100$ new spectral lines which were not observed before. The resolution of each of these peaks is about few MHz, making them suitable for laser frequency locking.

\end{abstract}

\maketitle

\section{Introduction}

Precision experiments invariably require calibration of device parameters using standards.
The most precisely measured physical parameter is frequency. To standardize frequency one
uses the energy gap between two electronic levels of an atom that has a narrow natural linewidth.
Even though the ground state hyperfine energy splitting of a $^{133}$Cs atom is currently set as
the atomic standard, it is likely to be replaced by optical transitions in atoms or ions~\cite{Udem02}.
In either case, the clock transition is a weakly coupled electronic transition at a short wavelength.
These clocks are based on laser cooled atomic samples. Similarly, ion trap or cold atom based quantum
information processors are also based on laser cooled atomic samples. Thus laser cooling has now become
a standard tool for any precision spectroscopy experiment. The lasers related to cooling or detection
are often required to be stable over long hours of experiment and in some cases like stable clock operation,
the laser needs to be stable for days. Here stability refers to long time drift of the laser frequency due
to thermal or mechanical stresses. These drifts even for the best laser designs can be as high as $10$'s MHz/hr
which is comparable to the cooling transition linewidth in commonly used atoms and ions. There are two ways to
mitigate this problem, namely, frequency locking the laser to an ultra-low expansion (ULE) optical cavity mode
or to an atomic  /molecular standard. The former demands good thermal and mechanical isolation of the cavity such
that the cavity length does not drift with time. In the best designs these drifts are about $10-100$~kHz/hr.
The less demanding option is to frequency lock the laser to an atomic or molecular reference spectral line.
However an optical cavity can be designed for any wavelength but it is not always possible to find an atomic or
molecular spectral line at a desired wavelength. One can in principle use the same atomic species as that of
the experiment in order to frequency lock the lasers. In case of multiple species of atoms or ions used in
an experiment, the setup becomes complicated due to the requirement of gas cells or hollow cathode lamps for each of the species. \\

Molecular electronic spectra usually covers a broad range of wavelengths, making them an ideal choice for
optical laser frequency standards which does not necessarily be accurate but has to be stable.
The most widely used molecules are iodine and tellurium, both are homo-nuclear and hence not IR active.
The electronic spectra of iodine shows strong absorption band starting at around $500~$nm for the
transition between its $X$ and $B$ electronic states which are the lowest two electronic states of the system~\cite{Geor89}.
Therefore, iodine lines are suitable to frequency lock lasers in the longer wavelength range. The tellurium spectra has been
observed starting from ultra violet (UV) wavelength of about $350~$nm onward. The main advantage of this molecule
is its large spin-orbit coupling and the presence of a large number of isotopes.
This combination allows a wide spectral range to be accessible in a single molecular specie~\cite{Verg82}.\\

Tellurium spectrum has been studied extensively in many contexts.
The most extensive work has been done by J. Cariou, P. Luc, Atlas du
spectre d’Absorption de la Molecule de Tellure, CNRS, Paris,
1980~\cite{Te2atlas:80}. Subsequent work has been done to extend
this atlas on both sides of the spectrum available in the
atlas~\cite{ Gil:91, You:94, Bar:85, Mct:90}. While longer wavelength region can be
made accessible by thermally exciting the molecule (having higher
temperature of the gas cell), short wavelength spectra are generated
by the strong spin-orbit coupling present in the molecule. The
progress made in identifying a plethora of spectral lines starting
in UV and continuing all the way above $600~$nm makes a good choice
as reference for quantum optics, quantum information and precision
measurement experiments~\cite{Raa:98, Wie:91, Coo:11, Dutta:16}. However,
the reference wavelength requirement of these experiments are
generally very specific and related to the atomic species being
probed. Therefore the reference spectral line should not be farther
than $1~$GHz away from the required frequency to be able to
frequency lock and shift by the use of standard acusto-optic
modulators (AOM). So far, the reported spectral lines in tellurium
are restricted in the $410-502$~nm region~\cite{ Ma:93, Sch:05, Bur:06} while the atlas also
includes some of the UV spectra~\cite{Makd:82}. However, the region
of interest for specific atoms are either the $470-490~$nm or closer
to the $450~$nm related to emission lines from barium and ytterbium
ions. Here we report experimentally measured more than $100$ new
lines close to $455~$nm wavelength and spreading over a spectral
range of $200~$GHz. The spectral resolution of these lines are about few MHz mainly limited by the natural linewidth. \\

\section{Experimental details}
\begin{figure}[t!]
\centering
\rotatebox{0}{\includegraphics*[width=\linewidth]{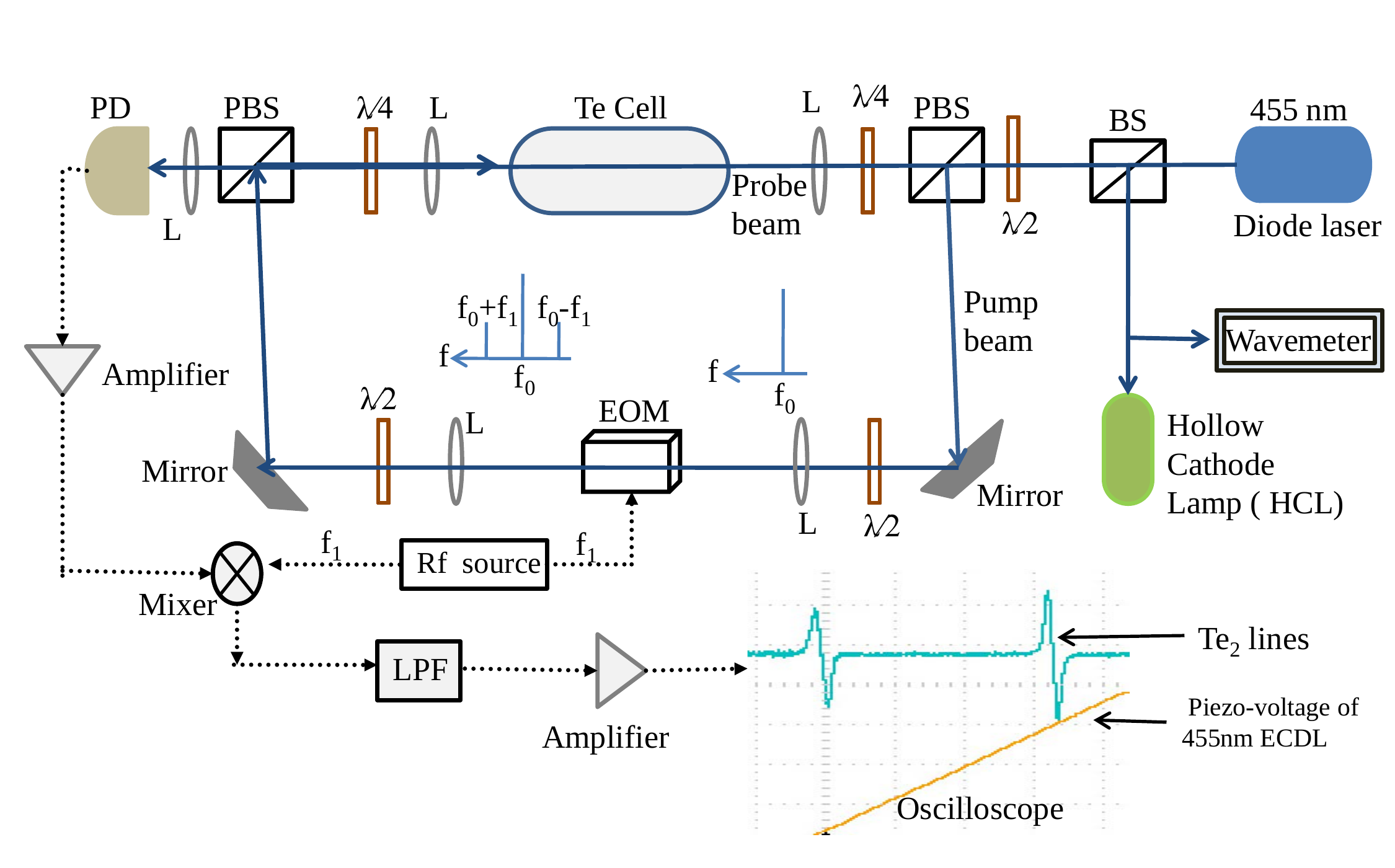}}
\caption{(Color online)Schematics of our setup for the MT spectroscopy where a saturated pump beam is sent through a tellurium vapour cell and a weak probe beam in the opposite direction. The beat signal, generated by this pump-probe beam after going through the Te cell, is picked up on the photodiode (PD) with a  responsivity of $0.64 $ W/A and then demodulated the signal by mixing the ac part of it with the modulation frequency (f$_1$) of  electro-optic modulator (EOM) which  has been employed to create the sidebands to the laser carrier frequency(f$_0$). The lineshape of the Te lines have been observed in the oscilloscope after doing the low-pass filtering (LPF) of the demodulated signal. The spectra thus, having a good signal to noise ratio that are free from any backgroud as shown in the schematic,  shows a potential for use in obtaining optical frequency referrence . PBS- polarizing beam splitter; BS- beam splitter; L- lens. }
\label{fig1}
\end{figure}

The experiment has been performed on a gas cell filled with
solid $^{130}$Te$_2$ in low pressure neon buffer gas. To obtain sufficient vapour pressure of tellurium the cell is heated up to $530^\circ$~C. Uniformly heated cell is a necessity to avoid solid deposit on the windows thus reducing the optical signal. A clamp heater has been used to heat the cell uniformly while the setup is mounted inside a heat shield which is then insulated using glass wool. A detailed thermal, mechanical and optical design of the cell is given in~\cite{Demu:16}. The experimental setup for the modulation transfer spectroscopy (MTS) is shown in figure~(\ref{fig1}). Modulation Transfer Spectroscopy has been chosen as opposed to saturation spectroscopy in order to avoid background noise in the spectral signal. In MTS two counter- propagating laser beams are used, one of which is modulated at a frequency of $\sim 5.8~$MHz. Due to the applied modulation on the pump beam, there are three frequencies of the pump and one frequency of the probe beam interacting simultaneously with the molecules. Thus a four-wave mixing mediated by the Tellurium molecules takes place inside the cell leading to a probe spectra which is a derivative of the saturation absorption resonance spectrum as shown in figure~(\ref{fig:4}). The line shape of each individual spectral line can be fitted using a known theoretical model to obtain the center frequency as well as the width of the spectral line. In order to scan over a wide range of frequencies yet having a good resolving power we have used external cavity diode laser~(ECDL) which is described in ref.~\cite{Dutta:16}. The laser line width is estimated to be below $1~$MHz with an output power of about $40$~mW. As shown in the figure~(\ref{fig1}), a part of the light is sampled out to simultaneously observe the wavelength in a Fizeau type wavemeter from {\it Toptica} while a second part probes a single Barium ion which is used as an absolute wavelength reference. The laser frequency can be scanned over $100~$GHz without any mode hop by scanning the voltage applied to the grating of the ECDL laser. However for the measurements reported here, the scan was performed piece-wise in order to maintain linearity of the piezo expansion as a function of the applied voltage.

\section{Measured spectral lines}

The spectral lines obtained in this experiment are tabulated in
Tab.~\ref{tab:Tspecs}. The relative frequencies are given
with respect to the line no. $82$ which was previously known spectral
line (line no. [1677]) from Atlas~\cite{Te2atlas:80}. The intensity
observed for each line is normalized to the line no. $82$, thus it is
referred to as relative intensity. The laser power varies as a
function of the frequency which has been normalized while calculating
the relative strength.
\begin{center}

\begin{longtable}{@{\extracolsep\fill}cccc}
\caption{New spectral lines of $^{130}$Te$_2$ molecule near $455.4~$nm wavelength. Relative frequency and amplitude of all the lines are also listed. The frequency  of line no. 82 is taken as the absolute frequency referrence. Relative amplitude has been defined as individual amplitude divided by the amplitude of line no. $82$. Comparing the Tellurium Atlas of Cariou and Luc~\cite{Te2atlas:80} twenty one lines have been identified, line no 82 is one of them.
\label{tab:Tspecs}} 
\\[-5pt]
 \hline 
 \textbf{Line no.} & \textbf{Wavenumber(cm$^{-1}$)}& \textbf{Rel. frequency(GHz)} & \textbf{Rel. strength}  \\ \hline 
\endfirsthead

\multicolumn{4}{c}%
{{\bfseries \tablename\ \thetable{} -- continued from previous page}} \\
\hline \textbf{Line no.} &\textbf{Wavenumber(cm$^{-1}$)}&
\textbf{Rel. frequency(GHz)} &
\textbf{Rel. strength} \\ \hline 
\endhead

\hline \multicolumn{4}{r}{\itshape Continued\dots} \\ 
\endfoot

\hline \hline
\endlastfoot

1   & 21948.5268  & 112.82  & 0.01  \\
2   & 21948.64822 & 109.18  & 0.14  \\
3   & 21948.68792 & 107.99  & 0.01  \\
4   & 21948.76464 & 105.69  & 0.01  \\
5   & 21948.85837 & 102.88  & 0.24  \\
6   & 21948.87204 & 102.47  & 0.91  \\
7   & 21948.97778 & 99.3    & 0.08  \\
8   & 21948.98812 & 98.99   & 0.02  \\
9   & 21949.00647 & 98.44   & 0.07  \\
10  & 21949.06317 & 96.74   & 0.29  \\
11  & 21949.09086 & 95.91   & 0.14  \\
12  & 21949.14123 & 94.4    & 0.03  \\
13  & 21949.15457 & 94      & 0.08  \\
14  & 21949.22462 & 91.9    & 0.04  \\
15  & 21949.25397 & 91.02   & 0.02  \\
16  & 21949.4301  & 85.74   & 0.11  \\
17  & 21949.44077 & 85.42   & 0.04  \\
18  & 21949.50114 & 83.61   & 0.04  \\
19  & 21949.50715 & 83.43   & 0.27  \\
20  & 21949.52049 & 83.03   & 1.56  \\
21  & 21949.62523 & 79.89   & 0.01  \\
22  & 21949.6836  & 78.14   & 0.01  \\
23  & 21949.91543 & 71.19   & 0.11  \\
24  & 21949.9201  & 71.05   & 0.03  \\
25  & 21949.93811 & 70.51   & 0.13  \\
26  & 21949.98615 & 69.07   & 0.33  \\
27  & 21950.1119  & 65.3    & 0.04  \\
28  & 21950.17895 & 63.29   & 0.27  \\
29  & 21950.22131 & 62.02   & 0.08  \\
30  & 21950.29403 & 59.84   & 0.18  \\
31  & 21950.33606 & 58.58   & 0.89  \\
32  & 21950.35507 & 58.01   & 0.01  \\
33  & 21950.37608 & 57.38   & 0.40  \\
34  & 21950.39743 & 56.74   & 1.11  \\
35  & 21950.41144 & 56.32   & 0.06  \\
36  & 21950.44046 & 55.45   & 0.04  \\
37  & 21950.50217 & 53.6    & 0.04  \\
38  & 21950.53986 & 52.47   & 0.01  \\
39  & 21950.59657 & 50.77   & 0.04  \\
40  & 21950.75868 & 45.91   & 0.01  \\
41  & 21950.76769 & 45.64   & 0.02  \\
42  & 21950.82673 & 43.87   & 0.08  \\
43  & 21950.87209 & 42.51   & 1.07  \\
44  & 21950.88744 & 42.05   & 0.01  \\
45  & 21950.89778 & 41.74   & 0.02  \\
46  & 21950.91212 & 41.31   & 0.10  \\
47  & 21950.9248  & 40.93   & 0.04  \\
48  & 21950.94815 & 40.23   & 0.01  \\
49  & 21950.97583 & 39.4    & 0.06  \\
50  & 21951.06356 & 36.77   & 0.02  \\
51  & 21951.19832 & 32.73   & 0.03  \\
52  & 21951.23601 & 31.6    & 0.04  \\
53  & 21951.26336 & 30.78   & 0.03  \\
54  & 21951.34609 & 28.3    & 0.18  \\
55  & 21951.37077 & 27.56   & 0.16  \\
56  & 21951.38111 & 27.25   & 0.02  \\
57  & 21951.38945 & 27      & 0.20  \\
58  & 21951.44182 & 25.43   & 0.91  \\
59  & 21951.46784 & 24.65   & 0.02  \\
60  & 21951.49319 & 23.89   & 0.01  \\
61  & 21951.54589 & 22.31   & 0.02  \\
62  & 21951.58859 & 21.03   & 0.07  \\
63  & 21951.62828 & 19.84   & 0.27  \\
64  & 21951.64663 & 19.29   & 0.03  \\
65  & 21951.67098 & 18.56   & 0.03  \\
66  & 21951.68933 & 18.01   & 0.02  \\
67  & 21951.702   & 17.63   & 0.67  \\
68  & 21951.73769 & 16.56   & 0.89  \\
69  & 21951.74103 & 16.46   & 0.02  \\
70  & 21951.82375 & 13.98   & 0.13  \\
71  & 21951.8411  & 13.46   & 0.01  \\
72  & 21951.93483 & 10.65   & 0.27  \\
73  & 21951.99053 & 8.98    & 0.06  \\
74  & 21952.04057 & 7.48    & 0.02  \\
75  & 21952.06192 & 6.84    & 0.02  \\
76  & 21952.12696 & 4.89    & 0.56  \\
77  & 21952.16799 & 3.66    & 0.44  \\
78  & 21952.17933 & 3.32    & 0.02  \\
79  & 21952.18467 & 3.16    & 0.01  \\
80  & 21952.2327  & 1.72    & 0.01  \\
81  & 21952.27373 & 0.49    & 0.02  \\
82  & 21952.29007 & 0       & 1.00  \\
83  & 21952.35212 & -1.86   & 0.02  \\
84  & 21952.3788  & -2.66   & 0.10  \\
85  & 21952.40115 & -3.33   & 0.16  \\
86  & 21952.40916 & -3.38   & 0.06  \\
87  & 21952.45519 & -4.91   & 0.01  \\
88  & 21952.51223 & -6.66   & 0.01  \\
89  & 21952.57827 & -8.64   & 0.01  \\
90  & 21952.63164 & -10.24  & 0.04  \\
91  & 21952.63965 & -10.48  & 0.27  \\
92  & 21952.70703 & -12.5   & 0.02  \\
93  & 21952.74372 & -13.6   & 0.01  \\
94  & 21952.8131  & -15.68  & 1.00  \\
95  & 21952.82011 & -15.89  & 0.16  \\
96  & 21952.93252 & -19.26  & 0.16  \\
97  & 21952.98022 & -20.69  & 0.02  \\
98  & 21953.01991 & -21.88  & 0.01  \\
99  & 21953.13032 & -25.19  & 0.01  \\
100  & 21953.16968 & -26.37  & 0.13  \\
101 & 21953.21838 & -27.83  & 0.01  \\
102 & 21953.23306 & -28.27  & 0.01  \\
103 & 21953.24907 & -28.75  & 0.01  \\
104 & 21953.36148 & -32.12  & 0.01  \\
105 & 21953.37282 & -32.46  & 0.02  \\
106 & 21953.39417 & -33.1   & 0.01  \\
107 & 21953.41852 & -33.83  & 0.04  \\
108 & 21953.44087 & -34.5   & 0.09  \\
109 & 21953.48257 & -35.75  & 0.03  \\
110 & 21953.54361 & -37.58  & 0.16  \\
111 & 21953.60131 & -39.31  & 0.01  \\
112 & 21953.61165 & -39.62  & 0.01  \\
113 & 21953.64868 & -40.73  & 1.33  \\
114 & 21953.69071 & -41.99  & 0.01  \\
115 & 21953.69705 & -42.18  & 0.01  \\
116 & 21953.70939 & -42.55  & 0.03  \\
117 & 21953.74041 & -43.48  & 0.01  \\
118 & 21953.74808 & -43.71  & 0.03  \\
119 & 21953.7791  & -44.64  & 0.19  \\
120 & 21953.83681 & -46.37  & 0.03  \\
121 & 21953.89819 & -48.21  & 0.01  \\
122 & 21953.93555 & -49.33  & 0.04  \\
123 & 21953.95089 & -49.79  & 0.20  \\
124 & 21953.95623 & -49.95  & 0.01  \\
125 & 21954.03161 & -52.21  & 0.02  \\
126 & 21954.04929 & -52.74  & 0.01  \\
127 & 21954.05496 & -52.91  & 0.04  \\
128 & 21954.08565 & -53.83  & 0.02  \\
129 & 21954.09232 & -54.03  & 0.01  \\
130 & 21954.12868 & -55.12  & 0.91  \\
131 & 21954.15136 & -55.8   & 0.03  \\
132 & 21954.16737 & -56.28  & 0.02  \\
133 & 21954.17471 & -56.5   & 0.02  \\
134 & 21954.18472 & -56.8   & 0.89  \\
135 & 21954.19906 & -57.23  & 0.01  \\
136 & 21954.20907 & -57.53  & 0.01  \\
137 & 21954.22875 & -58.12  & 0.07  \\
138 & 21954.28445 & -59.79  & 0.08  \\
139 & 21954.56598 & -68.23  & 0.11  \\
140 & 21954.57365 & -68.46  & 0.02  \\
141 & 21954.594   & -69.07  & 0.03  \\
142 & 21954.60434 & -69.38  & 0.03  \\
143 & 21954.61368 & -69.66  & 0.01  \\
144 & 21954.65438 & -70.88  & 0.36  \\
145 & 21954.69407 & -72.07  & 0.89  \\
146 & 21954.72042 & -72.86  & 0.01  \\
147 & 21954.75578 & -73.92  & 0.01  \\
148 & 21954.79481 & -75.09  & 0.04  \\
149 & 21954.80181 & -75.3   & 0.04  \\
150 & 21954.80515 & -75.4   & 0.01  \\
151 & 21954.8305  & -76.16  & 0.01  \\
152 & 21954.85752 & -76.97  & 0.19  \\
153 & 21954.89488 & -78.09  & 0.17  \\
154 & 21954.97326 & -80.44  & 0.02  \\
155 & 21955.02096 & -81.87  & 0.17  \\
156 & 21955.16506 & -86.19  & 0.78  \\
157 & 21955.19608 & -87.12  & 0.01  \\
158 & 21955.26713 & -89.25  & 0.01  \\
159 & 21955.27347 & -89.44  & 0.02  \\
160 & 21955.28415 & -89.76  & 0.03  \\
161 & 21955.37721 & -92.55  & 0.01  \\
162 & 21955.38588 & -92.81  & 0.00  \\
163 & 21955.39055 & -92.95  & 0.01  \\
164 & 21955.47328 & -95.43  & 0.01  \\
165 & 21955.53532 & -97.29  & 0.02  \\
166 & 21955.557   & -97.94  & 0.06  \\
167 & 21955.61537 & -99.69  & 0.02  \\
168 & 21955.68342 & -101.73 & 0.06  \\
169 & 21955.78583 & -104.8  & 0.56   \\  

\hline

\end{longtable}

\LTright=0pt
\LTleft=0pt
\end{center}

\section{Calibration of the frequencies}

All the Te$_2$ lines tabulated in Tab.~\ref{tab:Tspecs} have been observed by scanning the laser frequency over $100$~GHz on both side of the S$_{1/2}$-P$_{3/2}$ transition (455.4 nm) of barium ion. As mentioned before the whole range is scanned in piece-wise manner. Each piece of measurement is achieved by applying triangular voltage to the ECDL piezo at an amplitude $\sim 10-20~$V about a offset voltage at $20~$Hz frequency. The off-set voltage has been changed step-wise to cover the whole range of the frequency scan. The scale of the $100~$GHz frequency spectra has been obtained by stitching each neighborhood frequency range having more than one overlapping spectral lines. Within each scan range, about seven Te$_2$ lines have been observed~\cite{Dutta:16}. Even though the spectral lines are within $7~$GHz range, the peizo has been scanned by more than $10~$GHz such that non-linearity near the edge of the scan can be avoided. The piezo scanned voltage and the Te$_2$ lines,  both has common time axis and this time axis is converted to frequency using the conversion relation which is given by

\begin{equation}\label{eq:1}
f_1=f_0+\frac{\Delta f}{\Delta t} \Delta t,
\end{equation}

where $f_1$ is the unknown frequency and $f_0$ is the calibrated
frequency. The slope $\frac{\Delta f}{\Delta t}$ is the rate of
frequency scan as obtained from the linearity fit of scanned
voltage-time measurement and $\Delta t$ is the time difference
between the resonances $f_0$ and $f_1$
respectively. This linear transformation holds provided the piezo scanning
voltage is linear with time, otherwise non-linearity has to be taken
into account. Therefore the range of scanned voltage is
selected such that a linear fitting shows a $\chi^2 \sim
0.99$. The first piece of the scan range contains a well known barium resonance line S$_{1/2}$-P$_{3/2}$ which is used as $f_0$ as in eq.~(\ref{eq:1}). This frequency then acts as the absolute frequency for the frequency calibration. The frequency of the rest of the
Te$_2$ lines are determined by successive use of eq.~(\ref{eq:1}). Thus the uncertainty with which any of the resonance center frequencies can be measured is thus determined by the uncertainty of the rate of frequency scan, the accuracy in time with which each
resonance zero crossing can be determined, and the uncertainty of the frequency measurement $\delta f_ 0$. The uncertainty on the rate as obtained from the linearity as well as the uncertainty on the time measurement is negligible as compared with the $\delta f_ 0$, which is 30 MHz given by the resolution of the wavemeter.\\
In order to verify the correctness of this spectral stitching method, the frequency of almost all 
the lines  have been measured individually by actively locking the blue laser to the corresponding Te$_2$-line 
and these are compared with our calibrated data.  Even though, these lines are $100~$GHz 
away from the barium line, the frequencies match very well within the uncertainty dominated by the wavemeter uncertainty. The known Te-lines found in the
Atlas~\cite{Te2atlas:80} are compared with our measured data.
In particular, the line numbers $3$, $6 $, $15 $, $20 $, $31 $, $34  $, $43 $, $58 $, $67 $, $78 $, $82 $,  $94 $, $105 $, $113 $, $130 $, $134$, $ 145$, $ 154$ ,  $156$, $162$ and $169$ in the above table have been identified as the line numbers staring from $1667$ to $1687$ respectively in the Atlas data. The measured frequencies of  the line numbers $31$, $43$, $67$, $82$, $113$, $130$, $162$ and $169$  are found to be in good agreement with the Atlas data within the  wavemeter uncertainty whereas for the rest of the matched lines, the difference in frequency varies from $40-100$ MHz. 

\begin{figure}[t!]
\centering
\includegraphics[width=0.7\linewidth]{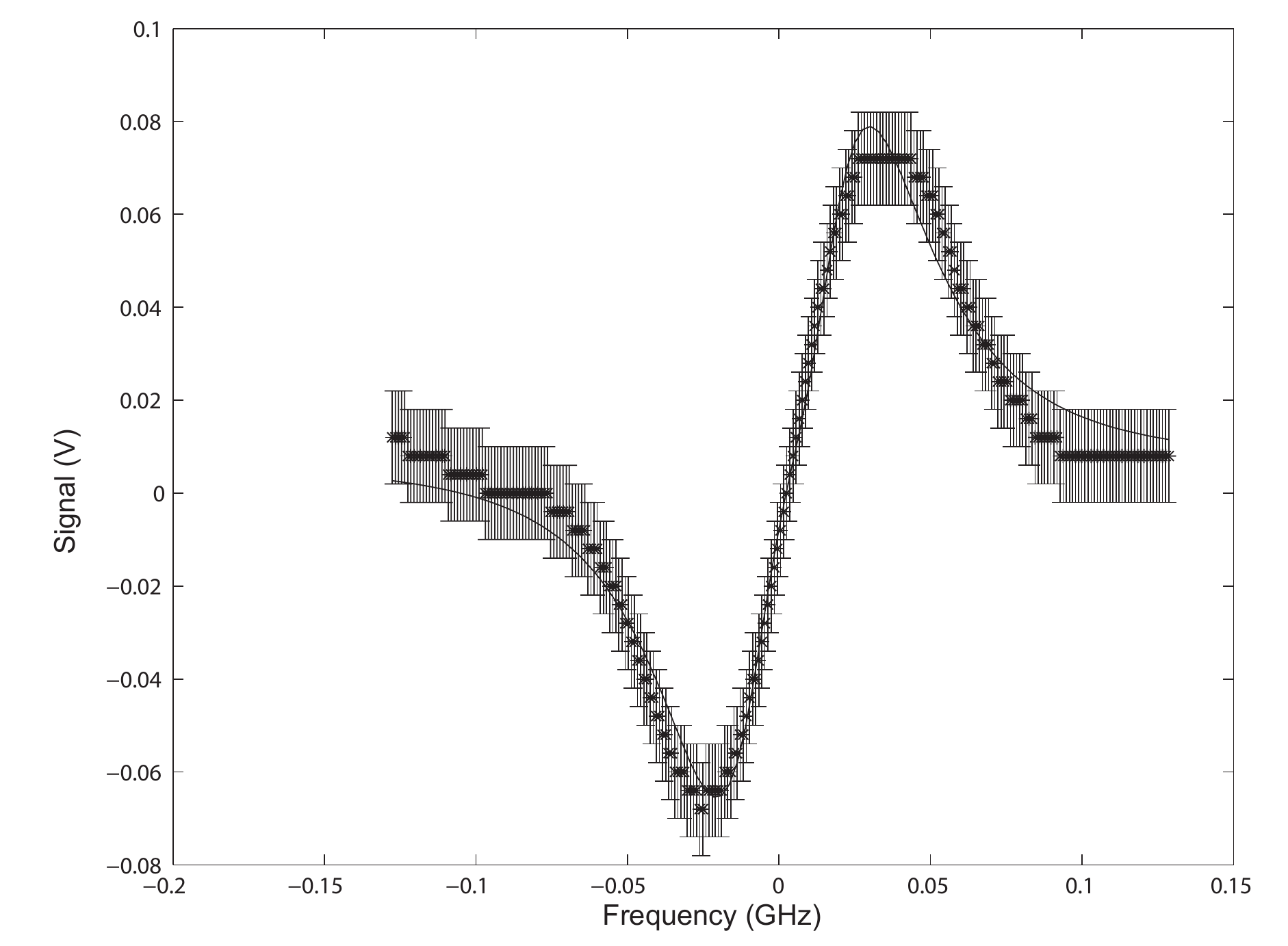}
\caption{The modulation transfer spectroscopy signal of a sample resonance line (line no. $83$ in the Table~\ref{tab:Tspecs}) as a function of the laser frequency. The $*$ denotes the experimental points along with the error bar while the solid line is a fit of the line-shape function with a reduced $\chi^2\sim 0.99$.}
\label{fig:4}
\end{figure}
\section{Line-shape and error determination}
The line center of individual resonance lines and their uncertainties has been obtained from the zero-signal crossing of the MTS spectrum. 
Figure~\ref{fig:4} shows one of the resonance spectrum fitted with a
line-shape function. The line shape of a MTS resonance is devoid of
any background slope unlike frequency modulation spectrum. This
shape can be described by a theoretical lineshape function as shown in \cite{line:82}.

The best fit provides the linewidth to be $20.9(4)~$MHz for the line shown in fig.~(\ref{fig:4}). Thus with an electronic suppression (about $100$) it is possible to frequency lock the laser with a bandwidth of a few hundred kHz for almost all lines shown in this work.

\section{Summary and discussion}

In this article we provide more than $100$ measured new resonance
lines of the Te$_2$ molecule beyond the known lines from the Atlas
with a frequency uncertainty of $30~$MHz over a range of $200~$GHz.
This is particularly important in terms of quantum optics,
communication and computing experiments as most of the cooling
lasers lies in the shorter wavelength range of the visible spectrum. In terms of laser frequency locking this range covers atomic resonance lines of atoms and ions including barium.


\begin{thebibliography}{999}

\bibitem{Te2atlas:80}
J. Cariou, P. Luc,\enquote{Atlas du Spectre d’Absorption de la
molKecule de Tellure, Laboratoire Aim´e–Cotton, CNRS II}, Orsay,
France (1980).

\bibitem{Udem02} Th. Udem, R. Holzwarth, T. W. H\"{a}nsch.\enquote{ Optical frequency metrology}, Nature 416 (2002) 233.
\bibitem{Geor89} Simon George, N. Krishnamurthy, \enquote{Absorption spectrum of iodine vapor - An experiment} American Journal of Physics 57, 850 (1989).
\bibitem{Verg82} J. Verg\`{e}s, C. Effantin, O. Babaky, J. d'Incan, S. J. Prosser, R. F. Barrow,\enquote{The Laser Induced Fluorescence Spectrum of Te2 Studied by Fourier Transform Spectrometry},  Phys. Scr. 25, 2 (1982).

\bibitem{Gil:91}
J. D. Gillaspy and Craig J. Sansonetti, \enquote{Absolute wavelength
determinations in molecular tellurium: new reference lines for
precision laser spectroscopy,} J. Opt. Soc. Am. B\textbf{8}
2414--2419 (1991)

\bibitem{You:94} Russell J. De Young, \enquote{Visible spectrum optical absorption in Te$_2$ vapor}, Appl. Phys. Lett. \textbf{64} 2631-2633 (1994).

\bibitem{Bar:85}
J. R. M. Barr, J. M. Girkin, A. I. Ferguson, G. P. Barwood, P.
Gill,W. R. C. Rowley, and R. C. Thompson, \enquote{Interferometric
frequency measurements of $^{130}$Te$_2$ transitions at 486 nm},
Opt. Commun. 54, 217-221 (1985).

\bibitem{Mct:90}
D. H. McIntyre, W. M. Fairbank, Jr. and S. A. Lee, T. W. Hansch, E.
Riis, \enquote{Interferometric frequency measurement of
$^{130}$Te$_2$ reference transitions at 486 nm}, Phys. Rev.
A\textbf{41}, 4632-4635 (1990).

\bibitem{Raa:98}
C. Raab, J. Bolle, H. Oberst, J. Eschner, F. Schmidt-Kaler, R.
Blatt, \enquote{Diode laser spectrometer at 493 nm for single
trapped Ba$^+$ ions}, Appl. Phys. B\textbf{67}, 683�1�7688 (1998).

\bibitem{Wie:91}
Carl E. Wieman, Leo Hollberg, \enquote{Using diode lasers for atomic
physics}, Rev. Sci. Instrum. \textbf{62}, 1-20 (1991).

\bibitem{Coo:11}
James Coker, Haoquan Fan, C. P. McRaven, P. M. Rupasinghe, T. Zh.
Yang, N. E. Shafer-Ray, and J. E. Furneaux, \enquote{Cavity
dispersion tuning spectroscopy of tellurium near 444.4 nm}, J. Opt.
Soc. Am. B\textbf{28} 2934-2939 (2011).

\bibitem{Dutta:16}
Tarun Dutta, Debashis De Munshi, and Manas Mukherjee,
\enquote{Absolute Te$_2$ reference for barium ion at 455.4nm} J. Opt.
Soc. Am. B \textbf{33}, 1177-1181 (2016).


\bibitem{Ma:93}
L.S. Ma, Ph. Courteille, G. Ritter, W. Neuhauser, R. Blatt,
\enquote{Spectroscopy of Te$_ 2$ with Modulation Transfer:Reference
Lines for Precision Spectroscopy in Yb$^+$ at 467 nm}, Appl. Phys.
B\textbf{57}, 159-162 (1993).

\bibitem{Sch:05}
T. J. Scholl, S. J. Rehse, R. A. Holt, and S. D. Rosner,
\enquote{Absolute wave-number measurements in 130Te2: reference
lines spanning the 420.9�1�7464.6-nm region}, J. Opt. Soc. Am.
B\textbf{22} 1128-1133 (2005).

\bibitem{Bur:06}
I.S. Burns, J. Hult, C. F. Kaminski, \enquote{Use of $^{130}$Te$_2$
for frequency referencing and active stabilisation of a violet
extended cavity diode laser}, Spectrochim. Acta A\textbf{63},
905-909 (2006).


\bibitem{Makd:82} Y. Makdisi, K S Bhatia,\enquote{Ultraviolet and visible spectra of tellurium I} J. Phys. B 15 (1982) 909.

\bibitem{Demu:16} D. De Munshi,\enquote{Precision Measurements to Explore
Underlying Geometries and Interactions
in a Trapped Ba+ Ion} Thesis, National University of Singapore, Singapore (2016).

\bibitem{line:82}
G. Camy, D. Pinaud, N. Courtier and Hu Chi Chuan,\enquote{Recent developments in high resolution saturation spectroscopy obtained by means of acousto-optic modulators} Rev. Phys. Appl.
\textbf{17} 357-363  (1982);

\end{thebibliography}
\end{document}